\documentclass[conference]{IEEEtran}
\IEEEoverridecommandlockouts
\usepackage{booktabs}
\usepackage{cite}
\usepackage{amsmath,amssymb,amsfonts}
\usepackage{algorithmic}
\usepackage{graphicx}
\usepackage{textcomp}
\usepackage{xcolor}
\usepackage{float}
\graphicspath{{./}{figures/}}

\def\BibTeX{{\rm B\kern-.05em{\sc i\kern-.025em b}\kern-.08em
    T\kern-.1667em\lower.7ex\hbox{E}\kern-.125emX}}
\begin{document}

\title{Learning Control of Neural Sound Effects Synthesis from Physically Inspired Models\\
}

\author{\IEEEauthorblockN{Yisu Zong}
\IEEEauthorblockA{\textit{Centre for Digital Music} \\
\textit{Queen Mary University of London}\\
London, United Kingdom \\
y.zong@qmul.ac.uk}
\and
\IEEEauthorblockN{Joshua Reiss}
\IEEEauthorblockA{\textit{Centre for Digital Music} \\
\textit{Queen Mary University of London}\\
London, United Kingdom \\
joshua.reiss@qmul.ac.uk}

}

\maketitle

\begin{abstract}
Sound effects model design commonly uses digital signal processing techniques with full control ability, but it is difficult to achieve realism within a limited number of parameters. Recently, neural sound effects synthesis methods have emerged as a promising approach for generating high-quality and realistic sounds, but the process of synthesizing the desired sound poses difficulties in terms of control. This paper presents a real-time neural synthesis model guided by a physically inspired model, enabling the generation of high-quality sounds while inheriting the control interface of the physically inspired model. We showcase the superior performance of our model in terms of sound quality and control.
\end{abstract}

\begin{IEEEkeywords}
Sound effects generation, Controllable sound synthesis, Physically inspired models
\end{IEEEkeywords}

\section{Introduction}
Sound effects play a crucial role in the field of sound design and production. Conventionally, the predominant approaches for their utilization are based on editing recorded audio. However, as there is an increasing demand for richer sound effects, the limitations of this time-consuming and constrained method have become increasingly apparent. Alternatively, modelling the physical phenomena of sound effects could provide a large number of variations based on the control parameters. Due to the complexity of modelling an entire physical environment, \emph{physically inspired models} are often preferred in practical implementations of procedural audio \cite{menexopoulos2023state}. This approach utilizes fundamental digital signal processing (DSP) components to perform simplified and approximate calculations of the physical system, incorporating both perceptually and physically meaningful controls that enable real-time generation. Given the trade-off between sound realism and computational complexity, they often incorporate numerous free parameters that pose challenges for optimization. A common way to improve the model performance is by exposing more parameters but at the expense of reduced ease of control \cite{zong2024machine}.

In recent years, data-driven neural sound synthesis has been the mainstream direction of academic research in sound synthesis, including Generative Adversarial Networks (GANs) \cite{Donahue2018AdversarialAS}, autoregressive models \cite{oord2016wavenet}, autoencoders \cite{caillon2021rave}, diffusion models \cite{chung2024t}, and show high potentials for generating realistic sounds. However, due to the limited interpretability of neural network models, controllable neural sound effects synthesis and its control mode remain an open question. One approach involves manipulating the output randomness \cite{andreu2022neural} or latent space \cite{wyse2022sound} of the model to obtain variations in target sounds; however, complete control over the generation direction is still elusive. Another common strategy is leveraging category labels associated with data, such as shoe type and ground surface for footstep sounds \cite{comunita2022neural} or emotions for knocking sounds \cite{barahona2020synthesising}. This method is constrained by the availability of labelled data and often relies on discrete labels only. Furthermore, directly controlling high-level audio features extracted from data could provide an intuitive control mode, such as loudness \cite{engel2019ddsp,barahona2024noisebandnet}, pitch \cite{engel2019ddsp} or other timbre features \cite{Nistal2020DrumGANSO,devis2023continuous}. However, this approach may not be optimal for explicit control since it does not directly reflect the physical process underlying sound generation.

The integration of control capabilities from physically inspired models with the generative potential of neural synthesis holds promise as a sound synthesis method. Physics priors could provide reliable and structured information to neural sound synthesis, e.g., ground reaction force curve of footstep sound \cite{serrano2023general}, or object interaction and resonance parameters of impact sound \cite{kamath2024example}, where the control parameters are generally extracted from well-defined physical equations. On the other hand, neural synthesis with the capacity to capture intricate sound details could serve as an auxiliary component within the system, enhancing the sound quality of physically inspired models without directly optimizing their complex inner structure. 

In this paper, we propose a neural sound effects synthesis system with an explicit control interface based on an example of a physically inspired explosion model. We first use synthesized sounds for training, and a latent discriminator is introduced to disentangle synthesized audio representations and the control behavior. Then, we compare two methods to perform the transfer to real sounds: supervised transfer using pseudo-label and an unsupervised transfer using CycleGAN \cite{zhu2017unpaired}. We conduct evaluations on both audio quality and control ability to demonstrate the effectiveness of our proposed method.

\section{Proposed Method}

\begin{figure*}[t]
    \centering
    \includegraphics[width=1.02\textwidth]{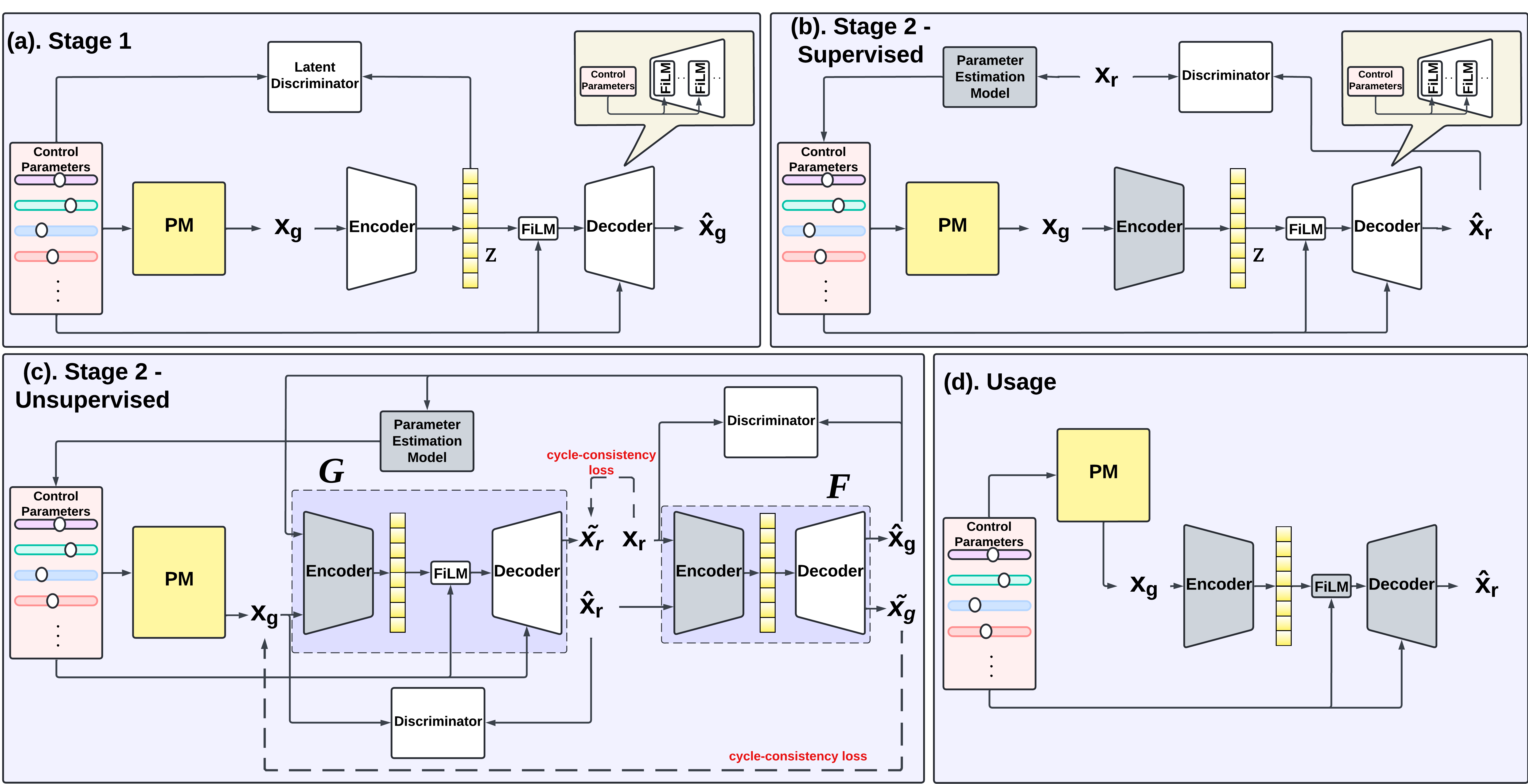}
    \caption{Flow diagram of proposed methods. Grey boxes represent frozen networks. (a). Representation learning stage for synthesized sounds \(x_g\) by the PM, achieving disentangled control facilitated by the latent discriminator. (b). Supervised transfer from \(x_g\) to real-world sounds \(x_r\) using pseudo-parameters obtained by a pre-trained parameter estimation model.  (c). Unsupervised transfer from \(x_g\) to \(x_r\) by CycleGAN. (d). Utilization of the proposed model. Control parameters and their corresponding \(x_g\) as inputs of the model to obtain \(x_r\).}
    \label{process}
\vspace{-1.3em}
\end{figure*}

\subsection{Explosion Model}
We use a physically inspired explosion model (PM) as an example in our experiment. The design concept is derived from the implementation by Andy Farnell \cite{farnell2010designing}, and the model can satisfy real-time queries\footnote{https://nemisindo.com/models/explosion}. The explosion sound is dominated by three main parts: rumble amount, air amount and dust amount, with eight continuous value control parameters: ``Rumble", ``Rumble Decay", ``Air", ``Air Decay", ``Dust", ``Dust Decay", ``Time Separation", and ``Grit Amount". Further details of this PM can be found in \cite{zong2024machine,farnell2010designing}.  

\subsection{Overall Architecture}
Considering the quality of synthesis and its applicability in real-time scenarios, we adopt a similar variational autoencoder (VAE) architecture to that of RAVE \cite{caillon2021rave}. It uses Pseudo Quadrature Mirror filters (PQMF) \cite{yu2019durian} to decompose the sound into multiple downsampled sub-signals, enabling real-time synthesis speed. The encoder is a convolutional downsampling network, and the decoder first uses convolutional upsampling layers and residual blocks, then waveform, loudness, and noise synthesizer networks are employed to process the signal. 

Our purpose is to synthesize real sounds \(x_r\) based on a set of continuous control parameters of the PM \(\theta_{x_g}\). To accomplish this, it is essential to acquire information about the controls of the PM and transfer them to real-world data. Therefore, we propose a two-stage training process: the first stage involves learning the latent representation and continuous control of PM for reconstruction, while the second stage focuses on translating generated sounds into real sounds.

\subsubsection{Learning Representation and Disentangled Control}
In the first stage, we train the VAE to reconstruct the generated sound by the PM along with its control parameters. For reconstruction, we aim to optimize a multi-resolution spectral loss \cite{engel2019ddsp} \(\mathcal{L}_s(x_g,\hat{x}_g)\),
\vspace{-0.2em}
\begin{align}
\hspace{-1.3cm}\mathcal{L}_s(x_g,\hat{x}_g)=\sum_{i\in N}(\|S(x_g)_i-S(\hat{x}_g)_i\|_1  \notag\\
&\hspace{-3cm} +\|\log S(x_g)-\log S(\hat{x}_g)_i\|_1)
\label{spectral}
\end{align}
\vspace{-1em}

where \(S(\cdot)\) is the magnitude spectrogram, and \(N\) is a set of Fast Fourier Transform sizes. The corresponding training objective \(\mathcal{L}_{\text{vae}}\) is derived from the Evidence Lower Bound (ELBO) \cite{DBLP:journals/corr/KingmaW13} as in RAVE \cite{caillon2021rave}.



To achieve disentangled control, the parameters should serve as additional inputs to the decoder. Instead of directly concatenating them with the latent vector \(z\), we utilize feature-wise linear modulation (FiLM) \cite{perez2018film} layers to inject the control information into the latent vector and decoder residual blocks.

However, the encoder may have already acquired sufficient information for reconstruction just from the data, leading to the decoder disregarding the control information. To address this issue, we employ a latent discriminator \cite{lample2017fader} that compels the encoder \(E\) to learn a representation without any control information. This discriminator \(D\) takes \(z\) as input and aims to output accurate control parameters \(\theta_{x_g}\), while the encoder aims to remove relevant information accordingly. 

The original latent discriminator \cite{lample2017fader} was designed for handling binary values; and in \cite{kawai2020attributes,devis2023continuous}, a multivariate discriminator was introduced to handle real values by partitioning the control parameter range into multiple equal segments and predicting the correct segment. In our case, our discriminator directly outputs the probability distributions for all the segments, so the loss of discriminator loss \(\mathcal{L}(D;E)\) and its corresponding generator loss for the encoder \(\mathcal{L}(E;D)\) are
\vspace{-0.2em}
\begin{gather}
\mathcal{L}(D;E)=-\mathbb{E}[log(p(\theta_{x_g}|z))]\\
\mathcal{L}(E;D)=-\mathbb{E}[log(1-p(\theta_{x_g}|z))]
\vspace{-1em}
\end{gather}
The total loss for our model at this stage is 
\begin{equation}
\mathcal{L}=\mathcal{L}_{\text{vae}}+\mathcal{L}(E;D)
\end{equation}

This stage is depicted in Figure \ref{process}(a). We train the model until the convergence of this loss, and subsequently freeze the encoder. Previous studies \cite{caillon2021rave,devis2023continuous} have incorporated an additional adversarial fine-tuning stage to enhance sound quality. However, in our case, we observed that satisfactory performance was achieved after stage 1, rendering adversarial training unnecessary.

\subsubsection{Transfer to Real Sound}
After completing stage 1 training, our model can be considered as a neural proxy of the PM model with disentangled controls. In the second stage, we further enhance sound realism by training our decoder on real sound data. The primary challenge in this transfer lies in the absence of ground-truth control parameters for real sounds due to misalignment between control parameters and any data labels. This discrepancy arises from our utilization of simulation-based control parameters rather than relying solely on strict physics. To address this issue, we experiment with both supervised and unsupervised learning modes: supervised learning using pseudo-label and unsupervised learning using CycleGAN \cite{zhu2017unpaired}.

\noindent\textit{Supervised learning using Pseudo-Label:} This stage is shown in Figure \ref{process}(b). To train our model in a supervised manner, paired data of \(x_r\) and \(x_g\) is required. We utilize a set of pseudo-parameters for \(x_r\) obtained through a sound matching task, i.e., estimating the PM control parameters of the best-matched generated sound for the real one. Following the approach in \cite{zong2024machine}, we train an end-to-end parameter estimation network. It should be noted that this method necessitates a differentiable implementation of the PM; therefore, applying it to non-differentiable models may require alternative derivative-free optimization methods, such as genetic algorithms. The pseudo-parameters \(\theta_{x_r}\) and their corresponding \(x_g\) are used as input to our model, with the reconstruction loss \(\mathcal{L}_s(x_r,\hat{x}_r)\) defined in equation (\ref{spectral}). 

Given our limited training dataset size,  we introduce a small random perturbation \(\delta\) to \(\theta_{x_r}\) during input processing to obtain \(\tilde{x_r}\), aiming to minimize the loss function \(\mathcal{L}_s(x_r,\tilde{x_r})\). Our assumption is that close parameters would generate similar sounds. Additionally, adversarial training is incorporated into this stage using MelGAN's discriminator \cite{kumar2019melgan}, where both its training objective \(\mathcal{L}_{adv}\) and the discriminator feature map loss \(\mathcal{L}_{FM}\) are employed. The total loss for this stage can be expressed as
\begin{equation}
\mathcal{L}=\mathcal{L}_s(x_r,\hat{x}_r)+\mathcal{L}_s(x_r,\Tilde{x}_r)+\mathcal{L}_{adv}+\mathcal{L}_{FM}
\end{equation}

\noindent\textit{Unsupervised learning using CycleGAN:}
CycleGAN \cite{zhu2017unpaired} offers an unsupervised approach for image-to-image translation without the need for paired data, and has been successfully applied to sound-to-sound tasks \cite{kaneko2019cyclegan,huang2019timbretron,yang2021unsupervised}. Following the method of CycleGAN, as illustrated in Figure \ref{process}(c), we train our model \(G\) using GAN framework to map input \(x_g\) onto \(x_r\), simultaneously training another original RAVE model \(F\) to map input \(x_r\) onto \(x_g\). The translation cycle should ensure the \textit{cycle-consistency}: mapping back from output space brings back to the original input space, i.e. \(F(G(x_g,\theta_{x_g}))=x_g\), and similarly for the reverse cycle: \(G(F(x_r),\theta_{\hat{x}_g})=x_r\), where \(\theta_{\hat{x}_g}\) is obtained by a parameter estimation network \cite{zong2024machine} exclusively trained on \(x_g\).  The cycle-consistency loss serves as the training objective:
\begin{equation}
\mathcal{L}_{cycle}=\|F(G(x_g,\theta_{x_g}))-x_g\|_1+\|G(F(x_r),\theta_{\hat{x}_g})-x_r\|_1
\end{equation}

We employ the identical GAN training objective as in the above supervised learning approach, thereby the total loss is
\begin{equation}
\mathcal{L}=\mathcal{L}_{cycle}+\mathcal{L}_{adv}(G)+\mathcal{L}_{adv}(F)
\end{equation}
\section{Experiments}
\subsection{Data}
For stage 1 training, we generated a dataset of 20,000 samples by randomly varying the parameter settings within the predefined range of the PM. All sounds are 3 seconds at the sample rate of 24 kHz. 

For the subsequent real data transfer stage, we curated 76 high-quality real explosion sound samples from Pro Sound Effects\footnote{https://www.prosoundeffects.com/hybrid-library/} and BBC Sound Effects\footnote{https://sound-effects.bbcrewind.co.uk/}. Our PM model is designed to generate a single explosion without including any environmental reflections or interactive effects such as glass shattering or secondary impacts caused by the explosion. Therefore, our data collection process adheres to the standard of avoiding obvious echoes and other interactive effects. Additionally, all collected samples were trimmed or zero-padded to maintain a consistent duration of 3 seconds at the sample rate of 24 kHz.

\subsection{Baselines}

To evaluate audio quality, we compared our supervised learning method (Supervised), CycleGAN method (Unsupervised), the original PM, and an enhanced version of the PM (PM-24params) \cite{zong2024machine}, wherein 24 parameters within PM were directly exposed.

\subsection{Evaluation Metrics}
\label{metrics}
We evaluated our model's generated sound quality and control ability. For sound quality evaluation, we adopted Fréchet Audio Distance (FAD) \cite{Kilgour2019FrchetAD}, Maximum Mean Discrepancy (MMD) \cite{gretton2012kernel}, and mel-cepstral distortion (MCD) \cite{kubichek1993mel}. 

To assess the control capability, we employed Spearman's rank correlation coefficient to evaluate the relationship between high-level audio features in the original PM outputs and our model outputs with identical control parameters. Specifically, we selected \textit{Boominess, Brightness, Roughness}, and \textit{Depth} from the Audio Commons project\footnote{https://audiocommons.github.io/} as our target audio features of interest.

\section{Results}
\subsection{Audio Quality}
We compared the sound matching results of PM (i.e. inputs of Supervised), PM-24params, and reconstruction quality of \textit{Supervised}. Also, we are interested in the overall audio quality with random parameters, since it could represent our model's reliability in generating realistic sounds. We compared quality with random parameters of PM, Supervised and Unsupervised across FAD and MMD as introduced in Section \ref{metrics}. Moreover, due to the limited size of our dataset, the pseudo-parameters derived from real data fail to encompass the entire parameter range. Consequently, we also evaluated the audio quality with random interpolated parameters within the pseudo parameter range of Supervised (Supervised (interpolation)). The results are shown in Table \ref{quality}, and we encourage readers to access our demo website\footnote{https://zys711.github.io/NeuralPM/} for subjective evaluation.

The Supervised method shows a major improvement compared with PM and PM-24params, indicating the neural network's efficiency in improving the design of a physically inspired model. For random control parameters, we observe that the Supervised (interpolation) has a pretty good performance, but this performance cannot extrapolate to the full parameter range, and the audio quality is even worse than PM. The Unsupervised method shows a stable performance in the entire parameter range but is slightly worse than the Supervised (interpolation). This is consistent with the expected since it can explore the parameter space more freely during training.

\subsection{Controls}
We compared the correlation between PM and Supervised, Supervised (interpolation), and Unsupervised, using randomly selected values for all control parameters with 100 samples. The results are shown in Table \ref{allcontrol}.
Additionally, we investigated the correlation when changing a single parameter while keeping all other parameters fixed. In this scenario, we present the results for Supervised (interpolation) in Table \ref{sup} and Unsupervised in Table \ref{unsup}.

The Supervised (interpolation) method shows the highest correlation with PM, yet it still lacks the ability to extrapolate across the entire parameter range. It demonstrates a significant positive correlation in terms of Roughness and Depth, while encountering challenges in capturing the characteristics of Boominess and Brightness. Similar trends are observed for the Unsupervised method, although there is an overall decrease compared to the Supervised (interpolation) approach. For single-parameter control, Supervised (interpolation) shows comparable performance to the overall correlations across all parameters. However, it is challenging for Unsupervised methods to replicate the same level of single-parameter control ability as PM. Without explicit labels, unsupervised learning has difficulty capturing the detailed relationships between individual parameters and specific sound characteristics, resulting in less effective control compared to the supervised approach.

\renewcommand{\arraystretch}{0.7}
\begin{table}[t]
\caption{Audio Quality Results}
\begin{center}
\begin{tabular}{c|ccc}
\toprule
 & FAD & MMD & MCD\\
\midrule
PM &29.29 & 119.54 & 1.20\\
PM-24params & 17.26 & 65.97&1.05 \\
Supervised  & \textbf{5.21} & \textbf{22.77}&\textbf{0.60}  \\
\midrule
PM (random) & 30.95 & 163.49&- \\
Supervised (interpolation)  & \textbf{8.87} & \textbf{58.35}&-  \\
Supervised (random)  & 37.76 & 190.09&-  \\
Unsupervised (random) & 12.71 & 95.54 & - \\

\bottomrule
\end{tabular}
\label{quality}
\end{center}
\vspace{-1em}
\end{table}

\begin{table}[H]
\caption{All parameters Control Correlations}
\vspace{-1.5em}
\begin{center}
\begin{tabular}{c|cccc}
\toprule
 & Boominess  &Brightness & Roughness & Depth\\
\midrule
Supervised (interpolation) & \textbf{0.70} &\textbf{0.66}& \textbf{0.86}&\textbf{0.95} \\
Supervised (random) & 0.03 & 0.18 & 0.52 &0.40 \\
Unsupervised (random) & 0.16 & 0.33 & 0.64 & 0.91 \\
\bottomrule
\end{tabular}
\label{allcontrol}
\end{center}
\end{table}
\vspace{-2em}

\begin{table}[H]
\caption{(Supervised(interpolation)) Single-Parameter control Correlations}
\begin{center}
\begin{tabular}{c|cccc}
\toprule
 & Boominess  &Brightness & Roughness & Depth\\
\midrule
Rumble & 0.96 &0.80& 0.71&0.88\\
Rumble Decay & 0.75 & 0.44 & 0.71 &0.73 \\
Air & 0.71 & 0.47 & 0.85 & 0.94 \\
Air Decay & 0.46 & 0.38 & 0.70 & 0.95 \\
Dust & 0.42 & 0.35 & 0.70 & 0.95 \\
Dust Decay & 0.44 & 0.34 & 0.70 & 0.90 \\
Time Separation & 0.38 & 0.52 & 0.59 & 0.88 \\
Grit Amount & 0.34 & 0.67 & 0.60 & 0.87 \\
\bottomrule
\end{tabular}
\label{sup}
\end{center}
\end{table}
\vspace{-2em}
\begin{table}[H]
\caption{(Unsupervised) Single-Parameter control Correlations}
\begin{center}
\begin{tabular}{c|cccc}
\toprule
 & Boominess  &Brightness & Roughness & Depth\\
\midrule
Rumble & -0.08 &-0.04& 0.78&0.93\\
Rumble Decay & -0.23 & -0.15 & 0.88 &0.27 \\
Air & 0.44 & 0.85 & -0.81 & -0.51 \\
Air Decay & 0.91 & 0.94 & -0.80 & 0.80 \\
Dust & 0.54 & 0.60 & -0.55 & -0.76 \\
Dust Decay & -0.37 & -0.31 & 0.81 & 0.90 \\
Time Separation  & -0.38 & 0.21 & 0.82 & 0.91 \\
Grit Amount & 0.44 & 0.28 & 0.35 & 0.86 \\
\bottomrule
\end{tabular}
\label{unsup}
\end{center}
\end{table}

\section{Conclusion}
We presented a real-time neural sound effects synthesis system that combines intuitive control from physically inspired models with the high-quality output of neural networks. The supervised method excels in terms of both quality and control within interpolated parameters, while the unsupervised method consistently delivers high audio quality performance across the entire parameter space at the expense of sacrificing fine-grained control. This integration of physically inspired models and neural networks offers a promising solution for achieving both control and realism in sound model design.

\clearpage

\bibliographystyle{IEEEtran}
\bibliography{IEEEabrv,IEEEexample}

\begin{thebibliography}{10}
\providecommand{\url}[1]{#1}
\csname url@samestyle\endcsname
\providecommand{\newblock}{\relax}
\providecommand{\bibinfo}[2]{#2}
\providecommand{\BIBentrySTDinterwordspacing}{\spaceskip=0pt\relax}
\providecommand{\BIBentryALTinterwordstretchfactor}{4}
\providecommand{\BIBentryALTinterwordspacing}{\spaceskip=\fontdimen2\font plus
\BIBentryALTinterwordstretchfactor\fontdimen3\font minus \fontdimen4\font\relax}
\providecommand{\BIBforeignlanguage}[2]{{%
\expandafter\ifx\csname l@#1\endcsname\relax
\typeout{** WARNING: IEEEtran.bst: No hyphenation pattern has been}%
\typeout{** loaded for the language `#1'. Using the pattern for}%
\typeout{** the default language instead.}%
\else
\language=\csname l@#1\endcsname
\fi
#2}}
\providecommand{\BIBdecl}{\relax}
\BIBdecl

\bibitem{menexopoulos2023state}
D.~Menexopoulos, P.~Pestana, and J.~Reiss, ``The state of the art in procedural audio,'' \emph{Journal of the Audio Engineering Society}, no.~12, pp. 826--848, 2023.

\bibitem{zong2024machine}
Y.~Zong, N.~Garcia-Sihuay, and J.~Reiss, ``A machine learning method to evaluate and improve sound effects synthesis model design,'' in \emph{Audio Engineering Society Conference: AES International Audio for Games Conference}, 2024.

\bibitem{Donahue2018AdversarialAS}
C.~Donahue, J.~McAuley, and M.~Puckette, ``Adversarial audio synthesis,'' in \emph{International Conference on Learning Representations}, 2018.

\bibitem{oord2016wavenet}
A.~v.~d. Oord \emph{et~al.}, ``Wavenet: A generative model for raw audio,'' \emph{arXiv preprint arXiv:1609.03499}, 2016.

\bibitem{caillon2021rave}
A.~Caillon and P.~Esling, ``Rave: A variational autoencoder for fast and high-quality neural audio synthesis,'' \emph{arXiv preprint arXiv:2111.05011}, 2021.

\bibitem{chung2024t}
Y.~Chung, J.~Lee, and J.~Nam, ``T-foley: A controllable waveform-domain diffusion model for temporal-event-guided foley sound synthesis,'' in \emph{IEEE International Conference on Acoustics, Speech and Signal Processing (ICASSP)}, 2024, pp. 6820--6824.

\bibitem{andreu2022neural}
S.~Andreu and M.~V. Aylagas, ``Neural synthesis of sound effects using flow-based deep generative models,'' in \emph{Proceedings of the AAAI Conference on Artificial Intelligence and Interactive Digital Entertainment}, vol.~18, no.~1, 2022, pp. 2--9.

\bibitem{wyse2022sound}
L.~Wyse, P.~Kamath, and C.~Gupta, ``Sound model factory: An integrated system architecture for generative audio modelling,'' in \emph{International Conference on Computational Intelligence in Music, Sound, Art and Design (Part of EvoStar)}, 2022, pp. 308--322.

\bibitem{comunita2022neural}
M.~Comunit{\`a}, H.~Phan, and J.~D. Reiss, ``Neural synthesis of footsteps sound effects with generative adversarial networks,'' in \emph{Audio Engineering Society Convention 152}, 2022.

\bibitem{barahona2020synthesising}
A.~Barahona-Rios and S.~Pauletto, ``Synthesising knocking sound effects using conditional wavegan,'' in \emph{SMC Sound and Music Computing Conference}, 2020.

\bibitem{engel2019ddsp}
J.~Engel, L.~Hantrakul, C.~Gu, and A.~Roberts, ``Ddsp: Differentiable digital signal processing,'' in \emph{International Conference on Learning Representations}, 2019.

\bibitem{barahona2024noisebandnet}
A.~Barahona-R{\'\i}os and T.~Collins, ``Noisebandnet: controllable time-varying neural synthesis of sound effects using filterbanks,'' \emph{IEEE/ACM Transactions on Audio, Speech, and Language Processing}, vol.~32, pp. 1573--1585, 2024.

\bibitem{Nistal2020DrumGANSO}
J.~Nistal, S.~Lattner, and G.~Richard, ``Drumgan: Synthesis of drum sounds with timbral feature conditioning using generative adversarial networks,'' in \emph{International Society for Music Information Retrieval Conference}, 2020.

\bibitem{devis2023continuous}
N.~Devis, N.~Demerl{\'e}, S.~Nabi, D.~Genova, and P.~Esling, ``Continuous descriptor-based control for deep audio synthesis,'' in \emph{IEEE International Conference on Acoustics, Speech and Signal Processing (ICASSP)}, 2023, pp. 1--5.

\bibitem{serrano2023general}
D.~Serrano and M.~Cartwright, ``A general framework for learning procedural audio models of environmental sounds,'' \emph{arXiv preprint arXiv:2303.02396}, 2023.

\bibitem{kamath2024example}
P.~Kamath, C.~Gupta, L.~Wyse, and S.~Nanayakkara, ``Example-based framework for perceptually guided audio texture generation,'' \emph{IEEE/ACM Transactions on Audio, Speech, and Language Processing}, 2024.

\bibitem{zhu2017unpaired}
J.-Y. Zhu, T.~Park, P.~Isola, and A.~A. Efros, ``Unpaired image-to-image translation using cycle-consistent adversarial networks,'' in \emph{Proceedings of the IEEE international conference on computer vision}, 2017, pp. 2223--2232.

\bibitem{farnell2010designing}
A.~Farnell, \emph{Designing sound}.\hskip 1em plus 0.5em minus 0.4em\relax Mit Press, 2010.

\bibitem{yu2019durian}
C.~Yu \emph{et~al.}, ``Durian: Duration informed attention network for multimodal synthesis,'' \emph{arXiv preprint arXiv:1909.01700}, 2019.

\bibitem{DBLP:journals/corr/KingmaW13}
D.~P. Kingma and M.~Welling, ``Auto-encoding variational bayes,'' in \emph{International Conference on Learning Representations}, 2014.

\bibitem{perez2018film}
E.~Perez, F.~Strub, H.~De~Vries, V.~Dumoulin, and A.~Courville, ``Film: Visual reasoning with a general conditioning layer,'' in \emph{Proceedings of the AAAI conference on artificial intelligence}, vol.~32, no.~1, 2018.

\bibitem{lample2017fader}
G.~Lample, N.~Zeghidour, N.~Usunier, A.~Bordes, L.~Denoyer, and M.~Ranzato, ``Fader networks: Manipulating images by sliding attributes,'' \emph{Advances in neural information processing systems}, vol.~30, 2017.

\bibitem{kawai2020attributes}
L.~Kawai, P.~Esling, and T.~Harada, ``Attributes-aware deep music transformation.'' in \emph{International Society for Music Information Retrieval Conference}, 2020, pp. 670--677.

\bibitem{kumar2019melgan}
K.~Kumar \emph{et~al.}, ``Melgan: Generative adversarial networks for conditional waveform synthesis,'' \emph{Advances in neural information processing systems}, vol.~32, 2019.

\bibitem{kaneko2019cyclegan}
T.~Kaneko, H.~Kameoka, K.~Tanaka, and N.~Hojo, ``Cyclegan-vc2: Improved cyclegan-based non-parallel voice conversion,'' in \emph{IEEE International Conference on Acoustics, Speech and Signal Processing (ICASSP)}, 2019, pp. 6820--6824.

\bibitem{huang2019timbretron}
S.~Huang, Q.~Li, C.~Anil, X.~Bao, S.~Oore, and R.~B. Grosse, ``Timbretron: A wavenet (cyclegan (cqt (audio))) pipeline for musical timbre transfer,'' in \emph{International Conference on Learning Representations}, 2019.

\bibitem{yang2021unsupervised}
J.~Yang, T.~Cinquin, and G.~S{\"o}r{\"o}s, ``Unsupervised musical timbre transfer for notification sounds,'' in \emph{IEEE International Conference on Acoustics, Speech and Signal Processing (ICASSP)}, 2021, pp. 3735--3739.

\bibitem{Kilgour2019FrchetAD}
K.~Kilgour, M.~Zuluaga, D.~Roblek, and M.~Sharifi, ``Fr{\'e}chet audio distance: A reference-free metric for evaluating music enhancement algorithms,'' in \emph{Interspeech}, 2019.

\bibitem{gretton2012kernel}
A.~Gretton, K.~M. Borgwardt, M.~J. Rasch, B.~Sch{\"o}lkopf, and A.~Smola, ``A kernel two-sample test,'' \emph{The Journal of Machine Learning Research}, vol.~13, no.~1, pp. 723--773, 2012.

\bibitem{kubichek1993mel}
R.~Kubichek, ``Mel-cepstral distance measure for objective speech quality assessment,'' in \emph{Proceedings of IEEE Pacific Rim Conference on Communications Computers and Signal Processing}, vol.~1, 1993, pp. 125--128.

\end{thebibliography}

\end{document}